\documentstyle[graphics]{mn}

\newcommand{\eg}{{\sl e.g. }}
\newcommand{\ie}{{\sl i.e. }}
\newcommand{\etal}{{\sl et al. }}

\newcommand{\Msun}{\mbox{$M_{\odot}$}}
\newcommand{\Rsun}{\mbox{R$_{\odot}$}}

\newcommand{\ltsimeq}{\raisebox{-0.6ex}{$\,\stackrel 
        {\raisebox{-.2ex}{$\textstyle <$}}{\sim}\,$}} 
\newcommand{\gtsimeq}{\raisebox{-0.6ex}{$\,\stackrel
        {\raisebox{-.2ex}{$\textstyle >$}}{\sim}\,$}}

\newcommand{\JHK}{$J\!H\!K~$}

\newcommand{\chis}{\mbox{$\chi^{2}~$}}

\newcommand{\kms}{$\,$km$\,$s$^{-1}$}
\newcommand{\gcm}{$\,$g$\,$cm$^{-3}$}
\newcommand{\ergs}{$\,$erg$\,$s$^{-1}$}

\title[IR spectroscopy of low-mass X-ray binaries II]
{Infrared spectroscopy of low-mass X-ray binaries II}
\author[R.M. Bandyopadhyay et al.]{
R.M. Bandyopadhyay,$^{1}$ T. Shahbaz,$^{1}$ P.A. Charles,$^{1}$ and T. Naylor$^{2}$ \\ 
$^{1}$University of Oxford, Department of Astrophysics, Nuclear Physics Building, Keble Road, Oxford, OX1 3RH, UK \\
$^{2}$Keele University, Department of Physics, Keele, Staffordshire, ST5 5BG, UK}

\begin{document}

\maketitle

\begin{abstract}

\noindent
Using CGS4 on UKIRT, we have obtained 2.00--2.45 $\mu$m infrared spectra of a number of low-mass X-ray binaries including Sco X-1, Sco X-2, and GX13+1.  Sco X-1 shows emission lines only, supporting our previous conclusion that the spectral type of the evolved secondary must be earlier than G5.  Emission lines are also seen in the spectrum of Sco X-2, confirming the identity of the IR counterpart.  We report the detection of CO bands in GX13+1 and estimate the most likely spectral type of the secondary to be K5$\sc iii$.  We also find P Cygni type profiles in the Brackett $\gamma$ lines of Sco X-1 and GX13+1, indicating the presence of high velocity outflows in these systems.  We present spectra of candidate IR counterparts to several other elusive X-ray binaries.  Finally, implications for the nature and classification of these systems are discussed.

\end{abstract}

\section{Introduction}
In low-mass X-ray binaries (LMXBs), a neutron star or black hole primary accretes material from its late-type companion star, producing intense X-ray emission.  LMXBs can be divided into subclasses  according to location within the Galaxy, accretion characteristics, and luminosity (\eg van Paradijs \& McClintock 1995).  The ``bright Galactic bulge'' sources (GBS) are located within $15^{\circ}$ longitude and $2^{\circ}$ latitude of the Galactic Centre (see \eg Warwick \etal 1988) and are among the most luminous X-ray sources in the Galaxy (typical $L_{X} \sim 10^{38}$ \ergs).  The GBS have shown no X-ray bursts and attempts to detect orbital variability have been unsuccessful, suggesting that their periods may be longer than those of canonical LMXBs (Charles \& Naylor 1992, hereafter CN92).  In addition, heavy obscuration in the Galactic Centre region has made optical study of the GBS nearly impossible.  There are as yet no observations which explain the dichotomy between the poorly understood GBS and the rest of the LMXBs.  The most likely theories suggest that the secondary stars in the GBS are late-type giants, in contrast to the quasi-main sequence stars in other LMXBs. 

However, the infrared provides us with an ideal window for observing these systems.  The IR has two primary advantages:  the late-type secondaries in LMXBs are brighter relative to the accretion discs, and, more importantly for the GBS, the ratio of $V$- to $K$-band extinction is nearly 10 (Naylor, Charles, \& Longmore 1991, hereafter NCL91).  Over the past eight years, we have developed a program of IR observations of X-ray binaries (XRBs), beginning with the discovery via colours or variability of candidates for the IR counterparts to the X-ray sources using precise X-ray and radio locations (NCL91).  Following this photometric survey we began an IR spectroscopic survey of LMXBs.  In 1995 we obtained IR spectra of the LMXB Sco X-1 and the GBS systems GX1+4 and GX13+1; these results were published in Bandyopadhyay \etal (1997), hereafter Paper I.

Continuing our spectroscopic survey of LMXBs, in this paper we present new results obtained from UKIRT in 1997.  In Paper I, we presented a \JHK spectrum of the prototype LMXB Sco X-1 taken in 1992 with an older, low resolution array and a short integration time.  We have now observed Sco X-1 at higher resolution and with a substantially longer exposure time.  We have also obtained the first $K$-band spectrum of the IR counterpart to the GBS Sco X-2 (GX 349+2), which was initially identified with a variable radio source (Cooke \& Ponman 1991); variability in the $R$-band counterpart was later found by Wachter \& Margon (1996).  In Paper I we presented a spectrum of the heavily obscured GBS GX13+1 which confirmed the identity of the IR counterpart.  We now present a new spectrum of this source which clearly shows the CO bands and metal lines of its late-type secondary.  Finally, we have obtained IR spectra of candidate stars within the X-ray/radio error circles of the GBS GX5-1, GX17+2, and the LMXB 4U2129+47.

\section{Observations and Data Reduction} 

We obtained $K$-band (2.00--2.45 $\mu$m) spectra using the Cooled Grating Spectrometer (CGS4) on the 3.8-m United Kingdom Infrared Telescope on Mauna Kea during the nights of 1997 July 1-3 UT.  The 75 l/mm grating was used with the 150 mm camera and the 256$\times$256 pixel InSb array.  Target observations were bracketed by observations of A-type stars for removal of telluric atmospheric features.  A journal of observations is presented in Table 1.

The standard procedure of oversampling was used to minimise the effects of bad pixels (Wright 1995).  The spectra were sampled over two pixels by mechanically shifting the array in 0.5 pixel steps in the dispersion direction, giving a full width half maximum resolution of 34 \AA\ ($\sim$460 \kms at 2.25 $\mu$m).  We employed the non-destructive readout mode of the detector in order to reduce the readout noise.  The slit width was 1.23 arcseconds which corresponds to 1 pixel on the detector.  In order to compensate for the fluctuating atmospheric emission lines we took relatively short exposures and nodded the telescope so that the object spectrum switched between two different spatial positions on the detector.  For some objects, the slit was rotated from its default north-south position to avoid contamination of the target spectrum by nearby stars.  Details of the design and use of CGS4 can be found in Mountain \etal (1990). 

\begin{table}
\caption{Journal of observations}
\begin{center}
\begin{tabular}{lccc}\hline
Object    &  Date      & UT    & Exposure time \\ 
          &            & (hrs) & (secs)       \\ \hline
Sco X-1   &  1/7/1997  &  6:55 & 3200   \\
Sco X-2   &  1/7/1997  &  8:35 & 2880   \\
	 		 &  1/7/1997  &  9:38 & 1920   \\
GX13+1    &  1/7/1997  & 10:48 & 2880    \\
          &  1/7/1997  & 11:55 & 2400  \\ 
GX17+2    &  2/7/1997  &  8:50 & 2400   \\
GX5-1$^{a}$ \#502 &  2/7/1997  & 10:10 & 960   \\
GX5-1$^{a}$ \#503 &  2/7/1997  & 10:49 & 960   \\
4U2129+47 &  3/7/1997  & 12:35 & 3360   \\ \hline
\end{tabular}
\end{center}
\noindent $^{a}$ Stars 502 and 503 are candidate stars within the X-ray error circle for GX5-1.\\
\end{table}	

The CGS4 data reduction system performs the initial reduction of the 2D images.  These steps include the application of the bad pixel mask, bias and dark subtraction, flat field division, interlacing integrations taken at different detector positions, and co-adding and subtracting the nodded images (see Daly \& Beard 1994).  Extraction of the 1D spectra, wavelength calibration, and removal of the telluric atmospheric features was then performed using IRAF.  A more detailed description of the data reduction procedure is provided in Paper I.

\section{Results}

\subsection{Sco X-1}

Our $K$-band spectrum of Sco X-1 is shown in Figure 1.  Strong emission lines of HI, HeI, and HeII are the dominant features; no absorption lines from the secondary star can be distinguished.  A list of identified lines is found in Table 2.

\begin{table}
\caption{Measured line wavelengths and equivalent widths}
\begin{center} 
\begin{tabular}{lcccc} \hline
Line          	&  $\lambda$   & Sco X-1 	& Sco X-2 	& GX13+1 \\ 
	      	& ($\mu$m) $^{a}$ & (\AA) 		& (\AA) 		& (\AA) \\ \hline
HeI	        	 &   2.059     & -2.8$\pm$0.4 & -26.0$\pm$1.6 & -3.7$\pm$0.6 \\
HeII 				 &	  2.098		& -1.0$\pm$0.2	&			&	\\
HeI 				 &   2.114 		& -1.0$\pm$0.2	&			&	\\
Br $\gamma$   	 &   2.165     & -12.6$\pm$0.5 & -29.0$\pm$1.8 & -4.8$\pm$0.4 \\
HeII 	   		 &   2.187     & -2.7$\pm$0.4 & 			&  \\
CaI (triplet)   &   2.262     &  	 		&  			&  1.2$\pm$0.5 \\ 
MgI	        	 &   2.283     &  	 		&  			&  0.8$\pm$0.4 \\ 
$^{12}$CO (2,0) &   2.295     &  	 		&  			&  5.8$\pm$0.6 \\
$^{12}$CO (3,1) &   2.324     &  	 		&  			&  3.2$\pm$0.6 \\ 
$^{13}$CO (2,0) &   2.345     &  	 		&   			&  0.6$\pm$0.3 \\
$^{12}$CO (4,2) &   2.353     &  	 		&  			&  3.7$\pm$0.6 \\ 
$^{13}$CO (3,1) &   2.377     &  	 		&  			&  1.6$\pm$0.5 \\
$^{12}$CO (5,3) &   2.384     &  	 		&  			&  3.8$\pm$0.7 \\ 
$^{12}$CO (6,4) &   2.416     &  	 		&  			&  1.6$\pm$0.7 \\ \hline
\end{tabular} 
\end{center}
\noindent $^{a}$ Typical wavelength calibration errors are approximately $\pm$0.001 $\mu$m. \\
\end{table}

\subsubsection{The secondary star}

In Paper I we presented a \JHK spectrum of Sco X-1, taken with the earlier 64x64 CGS4 array.  The spectrum showed no absorption features, and we concluded that the secondary is most likely a subgiant (see Paper I for a detailed discussion).  By modelling the appearance of two spectral template stars in the $K$-band at the distance and reddening of Sco X-1, we determined that the mass-donating star in Sco X-1 is of a spectral type which shows little or no CO features, \ie earlier than G5 for a subgiant (Kleinmann \& Hall 1986, hereafter KH86).  However, because the spectrum was taken with a short integration time on the older, lower resolution CGS4 array, there remained some doubt about this conclusion.

The Sco X-1 spectrum presented here removes this doubt.  With the new array and a $\sim$53 minute integration time, there is still no evidence of absorption features intrinsic to the secondary star in the $K$-band spectrum, only the emission lines expected from the disc and/or the heated face of the secondary.  The relatively long orbital period ($P$ = 18.9h) and the high mass transfer rate favour the formation of a large accretion disc (see Beall \etal 1984) and the optical spectrum is dominated by the disc (Schachter, Filippenko \& Kahn 1989); consequently we might expect to see emission from the disc in the IR.  However, based on our earlier modelling (see Paper I), in our new spectrum it is unlikely that contamination by the disc would be sufficiently high to completely obscure the strong CO features expected in the late-type subgiants.  Therefore the mass-donating star in Sco X-1 is of at least an early G-type, and perhaps even earlier, making Sco X-1 somewhat unusual.  The majority of identified secondaries in LMXB systems have been of a later type, most frequently K or M stars.  As such, it is interesting to note that Sco X-1, the ``prototypical LMXB'', may not be so typical after all.

\subsubsection{The P Cygni profile}

Despite the lack of absorption features from the secondary star, one absorption feature is present in our Sco X-1 spectrum.  The Br $\gamma$ emission line has a noticeable absorption dip on its blue edge: a P Cygni profile, indicative of a wind in the system.  By subtracting the measured velocity of the star from the velocity of the blue edge of the absorption feature, the maximum outflow velocity of the wind is obtained.  Although our spectrum has insufficient velocity resolution for an accurate measurement, we have made estimates of the radial velocity of Sco X-1 (using the HeI and HeII emission lines) and for the blue and red edges of the P Cygni profile; the results and the calculated outflow velocity are listed in Table 3.  A P Cygni profile has also been observed in the LMXB GX1+4; the $\sim$250\kms outflow in that system likely comes from the M5{\sc iii} donor star (Chakrabarty, van Kerkwijk, \& Larkin 1998).  However, the wind velocity we infer for Sco X-1, approximately 2600\kms, is supersonic, and thus unlikely to originate from the companion star (\eg Warner 1995, Dupree 1986).  As in CVs, the wind probably originates from the hot accretion disc, which produces a radiatively driven outflow.  Given the dissimilar sizes of the two systems ($P_{orb}\gtsimeq$260d for GX1+4; Chakrabarty \& Roche 1997) and the contrasting mechanisms for driving the wind, the order-of-magnitude velocity difference between the winds in GX1+4 and Sco X-1 is not unexpected.  Indeed, outflow velocities similar to that seen in Sco X-1 have been obtained in models of disc winds in CVs (\eg Mason \etal 1995).  However, to make accurate measurements of the outflow velocity and obtain information about the origin of the wind in Sco X-1 (\eg from the shape of the line profile), high resolution observations of the Br $\gamma$ line are required.

\begin{table}
\caption{P Cygni line profile parameters $^{a}$}
\begin{center}
\begin{tabular}{lcc}\hline
												&  Sco X-1			&  GX13+1       \\ \hline
Adopted rest velocity of star   		&  -230$\pm$40  	&  -70$\pm$30   \\
Red edge of Br $\gamma$ emission  	&  -420$\pm$30  	&  -430$\pm$30  \\
Blue edge of Br $\gamma$ absorption	&  -2830$\pm$50  	&  -2480$\pm$40 \\
Wind outflow velocity    				&  -2600$\pm$60  	&  -2400$\pm$50 \\ \hline
\end{tabular}
\end{center}
\noindent $^{a}$ All values given in \kms.  Errors quoted are formal errors only.\\
\end{table}	

\subsubsection{Shape of the continuum}

We have measured the slope of the $K$-band continuum of Sco X-1, after dereddening the spectrum using $E_{B-V}$ = 0.3 (determined from the 2200\AA\ interstellar absorption feature; Vrtilek \etal 1991).  We fit the data with a power law of the form $F_{\lambda} \propto \lambda^{\alpha}$ to represent the emission from a standard steady state accretion disc (\eg Frank, King, \& Raine 1992).  Using this model, we find a power law index $\alpha$= -3.17$\pm$0.02 for the IR spectrum of Sco X-1.  Shahbaz \etal (1996) found that the optical spectrum of Sco X-1 showed no evidence for any contribution from the secondary star and was well fitted with $\alpha$= -2.46; however, they used an older estimate of the reddening ($E_{B-V}$ = 0.15).  Using a reddening of 0.3 and normalizing the flux to the observed $V$ magnitude of Sco X-1 (van Paradijs 1995) by folding the spectrum through the filter response, we find $\alpha$ = -2.78 for the optical spectrum.  We then estimated the expected outer disc temperature ($T_{out}$) for a steady state disc with no irradiation and found $T_{out}\sim$4200 K.  The continuum slopes of the observed optical and IR spectra are different from that expected from a steady state disc at this temperature; these results are summarized in Table 4.  Figure 2 shows the observed optical and $K$-band spectra plotted with the simulated steady state disc for comparison.


\begin{table}
\caption{Continuum slopes: power law indices ($\alpha$) }
\begin{center}
\begin{tabular}{lcc}\hline
			&  Sco X-1			&  Steady State Disc   \\ 
			&	Observed			& $T_{out}$ = 4200 K		\\\hline
Optical  &  -2.78$\pm$0.09  &  -2.6$\pm$0.1   			\\
IR    	&  -3.17$\pm$0.06 &  -3.5$\pm$0.1  		 		\\\hline
\end{tabular}
\end{center}
\end{table}	

What is the cause of these discrepancies?  As Sco X-1 is a persistant source, we expect that the disc is irradiated and therefore hotter than a standard steady state disc; note that an irradiated disc is also required to fit the UV continuum observed by Vrtilek \etal (1991).  We therefore expect that the slope of the optical continuum will be considerably steeper (\ie bluer) than that predicted for a steady state disc.  Figure 2 illustrates that the observed optical spectrum is indeed much bluer than the simulated disc.  (Note that the simulated spectrum has been arbitrarily normalized; thus this discussion concerns only the shape of the continuum, {\it not} the flux level.)  However, it is unclear why the observed IR spectrum is shallower than the steady state disc model.  If the disc is irradiated, we would expect the slope of the IR spectrum to be steeper than a standard disc.  One possibility is that the secondary star contributes to the flux of the IR continuum.  A cool, late type (K/M) star would have a relatively shallow slope in the IR, whereas an earlier type (A/F) star would have a slope comparable to that expected from a standard disc.  In theory, therefore, flux from a cool secondary could cause the observed slope of the Sco X-1 spectrum to be shallower than expected from the disc alone.  However, if a late-type companion contributed sufficient flux to alter the slope of the continuum, that would imply a significant amount of the IR flux would be originating from the secondary star.  In this case, we would expect to see absorption features intrinsic to the secondary in the Sco X-1 spectrum (see \eg the modelling discussed in Paper I).  As there is no evidence for such features in our spectrum, it seems unlikely that flux from the mass-donating star is causing the discrepancy between the observed and expected continuum slopes.  It is therefore difficult to reconcile the continuum shape of the IR spectrum within the context of either a simple irradiated or steady state disc model.


The differences between the HeI and Br $\gamma$ line strengths in our 1992 and 1997 spectra may indicate variability in the level of X-ray irradiation of the disc.  After dereddening the 1992 $K$-band spectrum by $E_{B-V}$ = 0.3, we find the slope of the continuum to be -3.34$\pm$0.08, \ie somewhat steeper than the 1997 spectrum.  The bluer continuum and the stronger emission lines in the 1992 spectrum are consistent with a higher level of X-ray irradiation than in the 1997 data.  As we do not have any direct information about the X-ray state of Sco X-1 during the 1992 observations, it is difficult to compare the X-ray behaviour of the source during the two epochs quantitatively.  However, such changes in the continuum slope are consistent with the behaviour expected from a disc which is being irradiated by a variable level of X-ray emission.  Further, we note that the difference in the resolution of the two spectra and the highly changeable atmospheric absorption in the region of the 2.059$\mu$m HeI line makes any conclusions derived from the apparently large change between the HeI line strengths at the two epochs uncertain.

\subsection{Sco X-2}

Our $K$-band spectrum of Sco X-2 (GX349+2) was produced by combining our two observations for a total integration time of 80 minutes; the spectrum is presented in Figure 3.  The prominent features are the Brackett $\gamma$ and HeI emission lines from the accretion disc and/or heated face of the secondary.  The presence of these lines directly confirm the identification of the IR counterpart to the X-ray source.  The relatively low signal-to-noise (S/N $\sim$~8) despite the long exposure time is a result of the intrinsic faintness of the IR counterpart ($K\sim$~14.6).  The line identifications, measured wavelengths, and equivalent widths are listed in Table 2.

There is no evidence in our spectrum for the CO bandheads expected from a late-type secondary; however, the low S/N of our spectrum prevents us from drawing firm conclusions about the secondary's spectral type.  It is interesting to note that Southwell, Casares, \& Charles (1996; hereafter SCC96) found no evidence in optical spectra of Sco X-2 for the $\lambda$6495 Ca{\sc i}/Fe{\sc i} feature, which is a signature of late G- and K-type stars (Horne, Wade, \& Szkody 1985).  We also note our GX13+1 spectrum taken with CGS4 in 1995 {\it did} show clear evidence of the CO bandheads despite having a S/N comparable to that of our Sco X-2 spectrum shown here (Paper I).  To date, the secondary in Sco X-2 has not been spectroscopically detected in either optical or IR wavelengths.  The lack of spectroscopic features associated with the secondary may indicate that the mass-donating star in this system is of an earlier type than the K/M secondaries detected in a number of LMXBs.  Alternatively, the accretion disc contamination could be sufficiently high at both optical and IR wavelengths as to obscure late-type stellar spectral features in both our spectrum and the SCC96 optical spectrum.  

Several periodicities have been detected during observations of Sco X-2.  SCC96 suggested a period of $\sim$14d, whereas Wachter (1997) found a period of 22.5$\pm$0.1h.  As either of these periods could be the $\sim$1-day alias of the other, it is unclear which of the two is the true orbital period  (Barziv \etal 1997).  The 14d period is clearly inconsistent with a main sequence secondary star; assuming a Roche-lobe filling secondary, a spectral type of K0{\sc iii} would be required.  However, in this case we again note that we might expect to see the prominent CO absorption features of a K0{\sc iii} secondary in our Sco X-2 spectrum.  In contrast, a 22.5h orbital period would be insufficient to contain a late-type giant, but also indicates that the companion star is evolved; in this case, the secondary would most likely be a subgiant.  Several other XRBs with orbital periods $\sim$20h have subgiant secondaries, including Sco X-1, Cen X-4 ($P$=15.6h; Shahbaz \etal 1993), and Aql X-1 ($P$=19.1h; Shahbaz \etal 1996), suggesting that Sco X-2 may be similar.   


\subsection{GX13+1}

Figure 4 shows our $K$-band spectrum of GX13+1, produced by combining two spectra for a total on-source integration time of 88 min (S/N$\sim$~30).  The spectrum shows Br $\gamma$ and HeI emission, along with five $^{12}$CO bands and three $^{13}$CO bands which are characteristic of evolved late-type single stars (KH86).  Marginal detections of CaI and MgI are also present.  The line identifications, measured wavelengths, and equivalent widths are listed in Table 2.

\subsubsection{The secondary star}

In Paper I we reported the detection of Br $\gamma$ emission in GX13+1, which confirmed the identity of the IR counterpart to the X-ray source.  Also visible in the earlier spectrum were two CO bandheads; however, the much lower S/N of that spectrum only allowed us to constrain the secondary spectral type to be between K2 to M5.  Together with an estimate of the distance, we also determined that the secondary in this system must be a giant or subgiant.  These constraints were consistent with previous estimates of the secondary's spectral type (\eg Garcia \etal 1992).  Using our new spectrum of GX13+1, we have now estimated the spectral type of the companion star in GX13+1 by the technique of optimal subtraction, which minimizes the residuals after subtracting different template star spectra from the target spectrum.  This method is sensitive to the fractional contribution of the companion star to the total flux $f$, where 1-$f$ is the ``veiling factor'' (Marsh, Robinson, \& Wood 1994).

\begin{table}
\caption{Optimal subtraction of the GX13+1 companion star}
\begin{center} 
\begin{tabular}{lcc} \hline
Spectral Type  &  \chis (DOF=146)	& $f$	     \\ \hline
G5{\sc iii}	   &   495.0      		&  0.64$\pm$0.03  \\
G8{\sc iii}  	&   589.3      		&  0.75$\pm$0.04 \\
K0{\sc iii} 	&   421.3      		&  0.47$\pm$0.02 \\
K1{\sc iii}  	&   647.6      		&  0.66$\pm$0.04 \\ 
K4{\sc iii}	   &   436.5     			&  0.46$\pm$0.02 \\ 
K5{\sc iii} 	&   284.5    			&  0.45$\pm$0.02 \\
M0{\sc iii} 	&   446.6    			&  0.52$\pm$0.02 \\ 
M1{\sc iii} 	&   424.7    			&  0.41$\pm$0.02 \\
M2{\sc iii} 	&   481.3    			&  0.48$\pm$0.02 \\ 
M4{\sc iii} 	&   465.8    			&  0.35$\pm$0.02 \\
					&							&					  \\
G8{\sc iv}  	&   714.2    			&  0.82$\pm$0.07 \\ 
K0{\sc iv}  	&   507.6    			&  0.47$\pm$0.03 \\ 
K2{\sc iv}  	&   529.2    			&  0.44$\pm$0.02 \\ \hline
\end{tabular} 
\end{center}
\end{table}

First we determined the velocity shift of the spectrum of GX13+1 with respect to each template star spectrum by the method of cross-correlation (Tonry \& Davis 1979).  We then performed an optimal subtraction between each template star and the GX13+1 spectrum.  The optimal subtraction routine minimizes the residual scatter between the target and template spectra by adjusting a constant $f$, which represents the fraction of light contributed by the template star.  The scatter is measured by carrying out the subtraction and then computing the \chis between the resultant spectrum and a smoothed version of itself.  The constant $f$ is therefore the fraction of light arising from the secondary star.  The optimal values of $f$ are obtained by minimizing \chis.  

For this analysis we used a variety of giant and subgiant templates, ranging from G5 to M4.  The optimal subtraction was performed in the spectral range 2.28--2.39 $\mu$m in order to encompass the first four $^{12}$CO bands, which are the most prominent absorption features in the GX13+1 spectrum.  The results of this analysis are presented in Table 5.  The minimum \chis occurs for a K5{\sc iii} companion star, where the secondary contributes about 45\% of the flux at $\sim$2.3 $\mu$m.  Figure 5 illustrates the result of the optimal subtraction of the K5{\sc iii} template from the GX13+1 spectrum.  We note that the large IR variability which was observed by CN92 ruled out the possibility of the IR flux arising primarily from the mass-donating star; in other words, if the disc contribution to the IR flux was small, then we would expect the IR magnitude of GX13+1 to remain largely constant (as it does in GX1+4).  The substantial IR variability ($\sim$~1 mag at $K$) indicates that we should expect X-ray heating of the disc and secondary to be a major contributor to the IR flux from the system.  The result of our optimal subtraction, which indicates that the K5{\sc iii} secondary contributes less than half of the $K$-band flux, is completely consistent with this expectation.  


The ratio of the equivalent widths of $^{12}$CO to $^{13}$CO depends upon luminosity class (see \eg Dhillon \& Marsh 1995), ranging from 90 in main sequence stars to $\sim$10 in giants (Campbell \etal 1990).  To obtain a rough estimate of this ratio in GX13+1, we measured equivalent widths by first establishing the continuum as a linear function between marked points on either side of each feature.  The flux was then determined by summing the pixels within the marked area and subtracting the continuum.  Measurement of the ratio of the most clearly resolved $^{12}$CO and $^{13}$CO pair, the (2,0) bandheads, yielded a value of $\sim$9, which is comparable to that expected for a field giant and inconsistent with a main sequence secondary.

\subsubsection{The P Cygni profile}

In addition to the absorption features from the secondary, the Br $\gamma$ line in our GX13+1 spectrum exhibits a prominent P Cygni profile.  Similarly to Sco X-1, using the emission lines we have estimated the radial velocity of GX13+1 as well as measuring the location of the profile edges (see Table 3).  The inferred wind velocity, $\sim$~2400\kms, is similar to that calculated for Sco X-1; as in Sco X-1, the outflow probably originates in the disc rather than the mass-donating star.  A high resolution spectrum of the Br $\gamma$ line is necessary to obtain accurate information about the outflow in this system.  



\subsubsection{System parameters}

There have been several attempts at finding an orbital period in GX13+1.  Groot \etal (1996) have observed a possible 12.6$\pm$1 day modulation, while Corbet (1996) has reported a 25.2d period in the GX13+1 XTE ASM light curve.  We note that the Groot \etal period is half of that found by Corbet.  In contrast, Wachter (1996) found some evidence for a 19.5h periodicity in $K$-band photometric data.  Using Paczynski's formula for a Roche lobe filling star and an orbital period $P$~=~25.2d, the mean density for the secondary would be 0.0003\gcm; for a giant, this density corresponds to an approximate spectral type of K5.  The K5{\sc iii} spectral type we find for the secondary in GX13+1 therefore correlates well with the reported 25.2d period, and supports the possible identification of this period as an orbital modulation.  In addition, the low $L_{X}/L_{opt}$ ratio in GX13+1, possibly the result of having a large region involved in X-ray reprocessing (\ie a large disc), also reinforces the probability of a long orbital period in this system (CN92).  Our spectrum rules out an orbital period of 19.5h, which would be insufficient to contain the orbital separation of a 1.4\Msun ~neutron star and a Roche lobe filling K5{\sc iii} companion ($M\sim$~5\Msun, $R\sim$~25\Rsun; Allen 1973).  The nature of the 19.5h periodicity is therefore unclear.  

Using an apparent $K$ magnitude of 12 (CN92) and an absolute magnitude $M_{K}$ = -3.8 for a K5{\sc iii} star (Allen 1973; Koorneef 1983), together with a colour excess of $E_{B-V}$ = 5.7 (van Paradijs 1995), we find a distance to GX13+1 of 6.9 kpc.  There are several published estimates for the reddening ($A_{V}$) of GX13+1, ranging from $\sim$17.6 to $\sim$13.2 (see the discussion in CN92), leading to an uncertainty in our distance calculation of approximately 1 kpc.  Therefore we adopt a value of 7$\pm$1 kpc for the distance to GX13+1.  This value is consistent with the previous estimate of 8-9 kpc, which was based upon the mean distance to the Galactic Centre and has been used generally as an estimate for the distances to all of the GBS (see \eg Naylor \& Podsiadlowski 1993).

\subsubsection{Evolutionary status}

On the basis of their X-ray properties, LMXBs have been divided into two classes, known as $Z$ and atoll sources from the shape of their X-ray colour-colour diagrams (see van der Klis 1995 for a review).  GX13+1 has been classified as an atoll (less luminous) source (Hasinger \& van der Klis 1989, hereafter HK89), although earlier studies placed GX13+1 into the category with the $Z$ (high luminosity) sources (\eg White, Stella, \& Parmar 1988).  HK89 suggested that the evolutionary history of the $Z$ and atoll LMXBs could account for their observational differences, with the $Z$ sources having evolved secondaries and hence long orbital periods ($\gtsimeq$15h) whereas the atoll sources would have main sequence secondaries and $P_{orb}\ltsimeq$5h.  However, all of the proposed orbital periods found to date in GX13+1 are substantially longer than the $\sim$5h periods predicted for an atoll source and in fact are similar to the periods found in the $Z$ sources Sco X-1 and Cyg X-2.  Additionally, our spectrum shows unequivocally that the mass-donating companion in GX13+1 is an evolved late-type star, making a $\sim$5h orbital period impossible.  

There are several possibilities for resolving the discrepancy between the observed nature of GX13+1 and the predictions arising from the $Z$/atoll classification scheme.  First, GX13+1 may be mis-classified as an atoll source.  Both the high X-ray luminosity of GX13+1 and the detection of radio flux from the source are more typical of the luminous $Z$ LMXBs than atoll sources; additionally, a $\sim$60Hz QPO has recently been found (Homan \etal 1998).  However, its colour-colour diagram is not consistent with the X-ray properties of the six known $Z$ sources.  A second possibility is that the existence of ``hybrid'' systems such as GX13+1, GX9+9, and GX9+1, which exhibit both $Z$ and atoll characteristics, should be considered as a third ``intermediate'' class of objects where the $Z$ and atoll classes overlap (\eg Kuulkers \& van der Klis 1998).  Finally, it is possible that the evolutionary distinction between $Z$ and atoll LMXBs does not arise from a straightforward dichotomy between evolved and main sequence secondaries.  For example, some systems could have evolved companions but lower accretion rates and/or weaker magnetic fields than canonical $Z$ sources.  In this case, sources could exhibit atoll-type X-ray properties as well as long periods; in addition to GX13+1, the atoll source AC 211 ($P_{orb}$ = 17.1h) could fit into such a category (van der Klis 1992).  It would then become difficult to make predictions about the orbital period and nature of the companion star in a given system purely on the basis of its X-ray behaviour.  However, we note that the $Z$/atoll evolutionary dichotomy {\it could} still arise from fundamental differences in the types of mass-donating star in the two classes, but in a more specific manner than a simple division between evolved and main sequence companions.


\subsection{GX5-1, GX17+2, 4U2129+47}

In Figures 6 and 7 we show the spectra obtained of candidate IR counterparts to the GBS GX5-1 and GX17+2 and the LMXB 4U2129+47.  To date, no IR counterpart has been confirmed for any of these elusive LMXBs.  

\subsubsection{GX5-1}

Other than Sco X-1, GX5-1 is the brightest of the persistent LMXBs, but no X-ray periodicity has been found nor has an IR counterpart been identified (van der Klis \etal 1991).  An {\it Einstein} X-ray position and a precise radio position exist for GX5-1, but candidates for the IR counterpart to this heavily obscured source only become visible at $K$ (see \eg the finding charts in NCL91).  We obtained spectra of the two stars closest to the radio error circle, designated by NCL91 as stars 502 and 503.  Neither of the spectra, shown in Figure 6, show any evidence for the signature Br $\gamma$ emission we would expect in an LMXB.  Both stars 502 and 503 are faint, with $K$ magnitudes of $\sim$14 and $\sim$15 respectively (NCL91); combined with the relatively short integration time (16 min on each source), the generally poor quality of the two spectra (S/N $\sim$~3) is unsurprising.  However, in such a bright X-ray source, we might expect the emission lines from the disc to be especially strong; while not conclusive, our spectra therefore cast doubt on the potential of either of these two stars to be the IR counterpart of the X-ray source.  We note that there is a star northeast of star 502 (designated as star 513 by NCL91) which also lies within the radio error circle of GX5-1 and as such is also a prime candidate for the IR counterpart.  Due to the star's extreme faintness ($K\ltsimeq16$), we did not obtain a spectrum for this candidate.  Obtaining photometric and spectroscopic information on star 513 should therefore be a primary objective for future IR observations of GX5-1.

\subsubsection{GX17+2}

On the basis of its X-ray position, the GBS GX17+2 was optically identified with a G star known as ``star TR'' more than 25 years ago (Tarenghi \& Reina 1972; Davidsen, Malina, \& Bowyer 1976).  The location of GX17+2 was further refined by a subarcsecond radio position (Hjellming 1978) but the putative counterpart has shown almost no optical variability (\eg Margon 1978).  We note, however, that Deutsch \etal (1996) have found a small discrepancy between the optical and radio positions of GX17+2.  The IR counterpart to star TR ($K\sim$14) has shown variability, but a consistent fit for the colours, extinction, distance, and spectral type could not be found (NCL91).  Based on the variety of X-ray periods which have been suggested for this source, Bailyn \& Grindlay (1987) theorized that GX17+2 is a triple system, with a giant star in orbit around a short-period LMXB.  However, the discrepancy between the extinctions derived from optical and X-ray measurements led NCL91 to conclude that star TR is a foreground star unassociated with the X-ray source which is superimposed upon a highly reddened IR counterpart to GX17+2.  Our $K$-band spectrum, shown in the top panel of Figure 7, shows no evidence either for emission lines or absorption features from a late-type IR counterpart obscured by star TR.  Indeed, the featureless spectrum is consistent with an early G-type evolved star or a main sequence star earlier than G5 (KH86), as would be expected from star TR if it is an unassociated, foreground G star.  Therefore, our spectrum has too short an integration time and/or insufficient resolution to show features from a faint, variable IR counterpart to GX17+2 which may be partially hidden behind star TR.  

\subsubsection{4U2129+47}

The neutron star LMXB 4U2129+47 is known to have an orbital period of 5.2h and has an optical counterpart, V1727 Cyg (Thorstensen \etal 1979).  Radial velocity studies produced estimates for the masses of the compact object and secondary star of 0.6$\pm$0.2\Msun ~and 0.4$\pm$0.2\Msun ~respectively (Thorstensen \& Charles 1982).  Then, after entering X-ray quiescence in 1983, all evidence of photometric variability and radial velocity variations disappeared (\eg Garcia \etal 1989, hereafter G89).  The optical counterpart visible during the low state is an F8{\sc iv} (Cowley \& Schmidtke 1990) whose colours were initially observed to be inconsistent with a normal star (Kaluzny 1988).  However, subsequent observations failed to confirm this abnormality in the optical colours of the F star (Deutsch \etal 1996).  G89 proposed that 4U2129+47 is a triple system, although it has also been suggested that the F star is a foreground source unassociated with the system.  Our $K$-band spectrum of the F star appears in the bottom panel of Figure 7.  The spectrum is featureless (the spike at $\sim$2.05 $\mu$m is a residual of the telluric absorption feature removal); we see no indication of emission from an accretion disc, nor do we see any evidence for absorption features from the M{\sc v} star suggested by G89 as a hypothetical third member of the triple system.  However, we note that even in the $K$-band, the flux from a faint M{\sc v} star would likely be obscured by the brighter F star.


\section{Conclusions}

We have presented $K$-band spectra of the LMXBs Sco X-1, Sco X-2, and GX13+1.  The IR spectrum of Sco X-1 exhibits the emission features expected from a luminous accretion disc but does not show any absorption features arising from the system secondary, reinforcing our conclusion that the secondary is a subgiant of spectral type G5 or earlier.  The spectrum of the proposed counterpart to Sco X-2 shows Br $\gamma$ emission from the accretion disc, confirming the identification.  The $K$-band spectrum of GX13+1 exhibits both Br $\gamma$ emission and CO absorption bands; optimal subtraction indicates that the secondary in GX13+1 is most likely a K5{\sc iii}, with the accretion disc contributing $\sim$50\% of the $K$-band flux.  In addition, the Br $\gamma$ emission lines in both Sco X-1 and GX13+1 show P Cygni profiles, indicating the presence of outflows in both systems with velocities $\sim$2000\kms.  Spectra of two stars in the error circle of GX5-1 do not show the Br $\gamma$ emission signature which would identify the IR counterpart.  Similarly, spectra of the G star at the position of GX17+2 and the F star at the position of 4U2129+47 fail to exhibit any evidence that they are physically associated with the X-ray sources.  

It is interesting to note that while our GX13+1 spectrum clearly shows the features of the K giant secondary, neither Sco X-1 nor Sco X-2 appear to exhibit any spectral lines characteristic of late-type (late G to M) stars.  Significant similarities in the behaviour of Sco X-1 and Sco X-2 have also been seen in the optical and X-rays, indicating a fundamental similarity in the nature of the two systems.  Furthermore, the clear differences in these spectra occur {\it despite} the fact that GX13+1, Sco X-1, and Sco X-2 all have relatively long orbital periods and are luminous, persistent XRBs.  Far from supporting the idea that most of the GBS have late-type evolved secondaries, this result leads us to speculate that the mass-donating stars in Sco X-1 and Sco X-2 may prove to be very different than previously assumed.  

There have been two primary schemes proposed for the classification of LMXBs.  In a statistical study of a flux-limited sample of LMXBs, Naylor and Podsiadlowski (1993, hereafter NP93) asserted that the GBS distribution is associated with that of M giants in the Galactic bulge.  They classified Sco X-1, Sco X-2, and GX13+1 (along with seven other XRBs) as bulge sources on the basis of their location, the characteristics of their X-ray spectra, and the fact that they are persistent X-ray sources.  In contrast, the $Z$/atoll scheme categorizes LMXBs based on the shape of their X-ray colour-colour diagrams (HK89).  With this method, six of the sources classified as GBS by NP93, including Sco X-1 and Sco X-2, are labelled as $Z$-sources, whereas the remaining GBS, including GX13+1, are placed in the atoll category.  Note that the $Z$ and atoll sources appear to be randomly scattered amongst the bulge and disc populations categorized by NP93.  HK89 suggested that a primary difference between the $Z$ and atoll sources may originate with the mass-donating stars; namely, that the $Z$ sources have evolved secondaries while the atoll sources have main sequence companions.  It is clear from our spectra that the dichotomy between $Z$ and atoll sources cannot be this straightforward, as both Sco X-1 and GX13+1 have evolved secondaries.  However, if the spectral type of the secondaries in Sco X-1 and Sco X-2 prove to be considerably earlier than the K giant found in GX13+1, then the evolutionary scenarios of Sco X-1 and Sco X-2 are indeed likely to be quite distinct from sources such as GX13+1.  We also note that the $Z$ source Cyg X-2 has an evolved A9 companion (Casares, Charles \& Kuulkers 1998); Sco X-1 also has an evolved secondary of type earlier than G5, and Sco X-2 may be similar.  

We therefore hypothesize that the secondaries in the six $Z$ sources may all be {\it early-type evolved stars}.  We speculate further that the distinction between the evolutionary history of the $Z$ and atoll sources may not be that the former have evolved secondaries and the latter have main sequence companions, but instead that $Z$ sources have {\it evolved early-type} companions while most other LMXBs have {\it late type} (either evolved or main sequence) secondaries.  If this were the case, it may also help to explain why the IR counterparts for the other three $Z$ sources (GX5-1, GX17+2, and GX340+0) have proven so elusive (although GX17+2 is a special case due to the foreground star).  Until now it has been assumed that the secondaries in these systems would be late-type evolved stars and therefore bright in the $K$ band ($V-K \sim$~4).  Yet the few candidate stars found in deep $K$-band images of the GX5-1 and GX340+0 fields have failed to show photometric or spectroscopic evidence that they are in fact the counterparts.  To elude detection, a late-type evolved counterpart would have to be at a greater distance than we believe from estimates of the X-ray column densities.  However, if the secondaries in these systems are early-type stars (with $V-K \sim$~0), then the counterparts would naturally be fairly faint in the $K$-band.  It would then be unnecessary to invoke a significantly larger extinction and/or distance to these sources in order to explain our inability to locate their IR counterparts.  

Our spectra provide clear evidence that neither the bulge/disc nor the $Z$/atoll schemes can be explained by a simple distinction between evolved and main sequence secondary stars.  However, we note that the early-type/late-type dichotomy proposed here {\it could} fit into the $Z$/atoll classification scheme {\it without} having to disregard the bulge/disc association made by NP93.  Although currently the evidence for this claim is limited, we believe that this hypothesis is well worth exploring.





\section{Acknowledgements}

The authors would like to thank Deepto Chakrabarty for helpful discussions about winds in X-ray binaries, and Erik Kuulkers for answering our questions about $Z$ and atoll sources.  We would also like to thank Tom Marsh for the use of his $\sc molly$ routine.  The data reduction was carried out using the $\sc iraf$ and $\sc ark$ software packages on the Oxford {\sc starlink} node.  TN was supported by a PPARC Advanced Fellowship.  The United Kingdom Infrared Telescope is operated by the Royal Observatory Edinburgh on behalf of the UK Particle Physics and Astronomy Research Council.

\clearpage
\newpage
\clearpage

\begin{figure} 
\begin{center}
\rotatebox{90}{
\resizebox{0.5\textwidth}{0.7\textheight}{\includegraphics{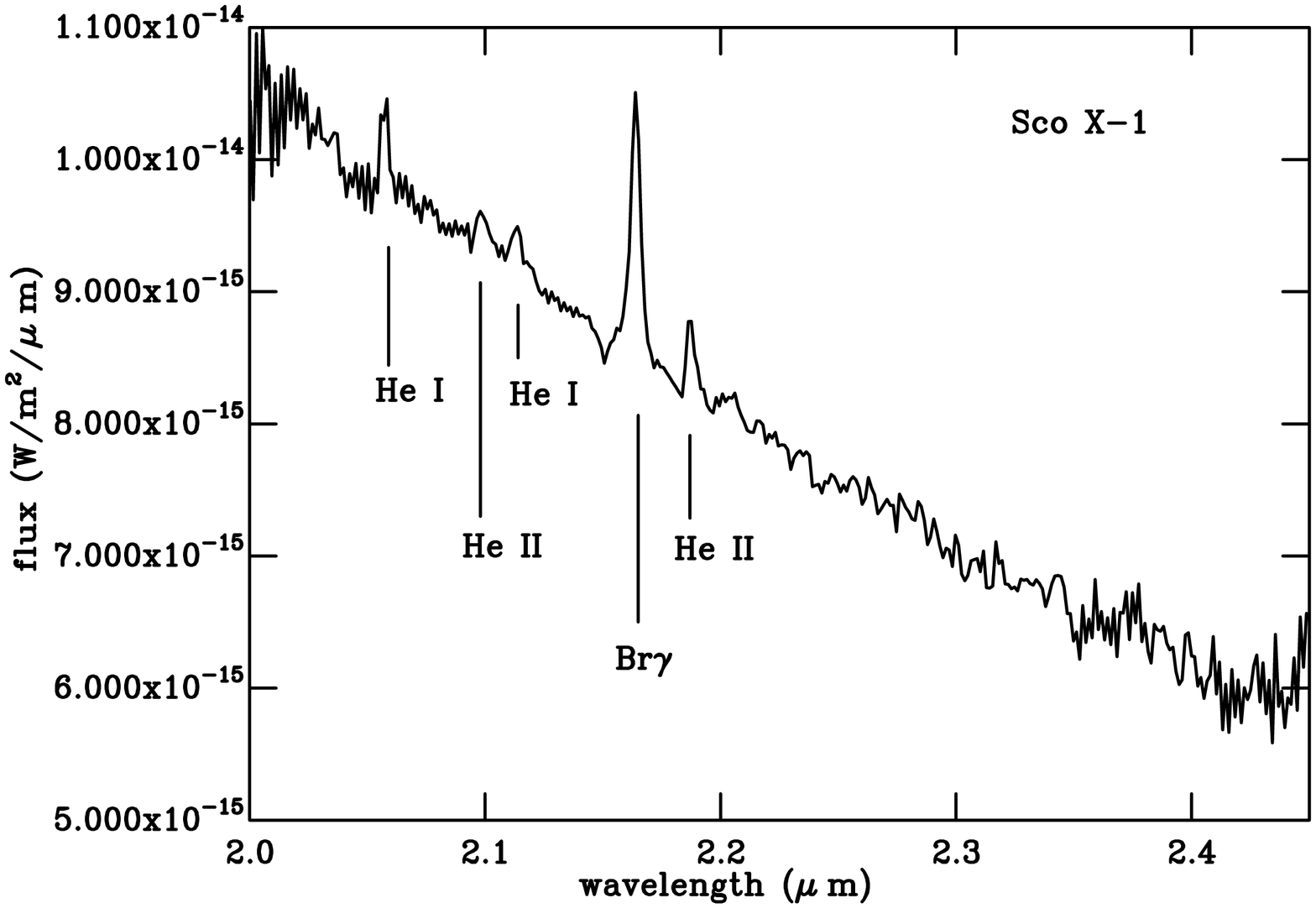}}}
\end{center}
\onecolumn{\caption{ $K$-band spectrum of Sco X-1 ($K\simeq$~11.8). Strong emission lines of He I, He II, and Br $\gamma$ are evident, but no absorption features from the secondary can be distinguished.  Note that the Br $\gamma$ line has a P Cygni profile.}}
\end{figure} 

\begin{figure} 
\begin{center}
\rotatebox{-90}{
\resizebox{0.5\textwidth}{0.7\textheight}{\includegraphics{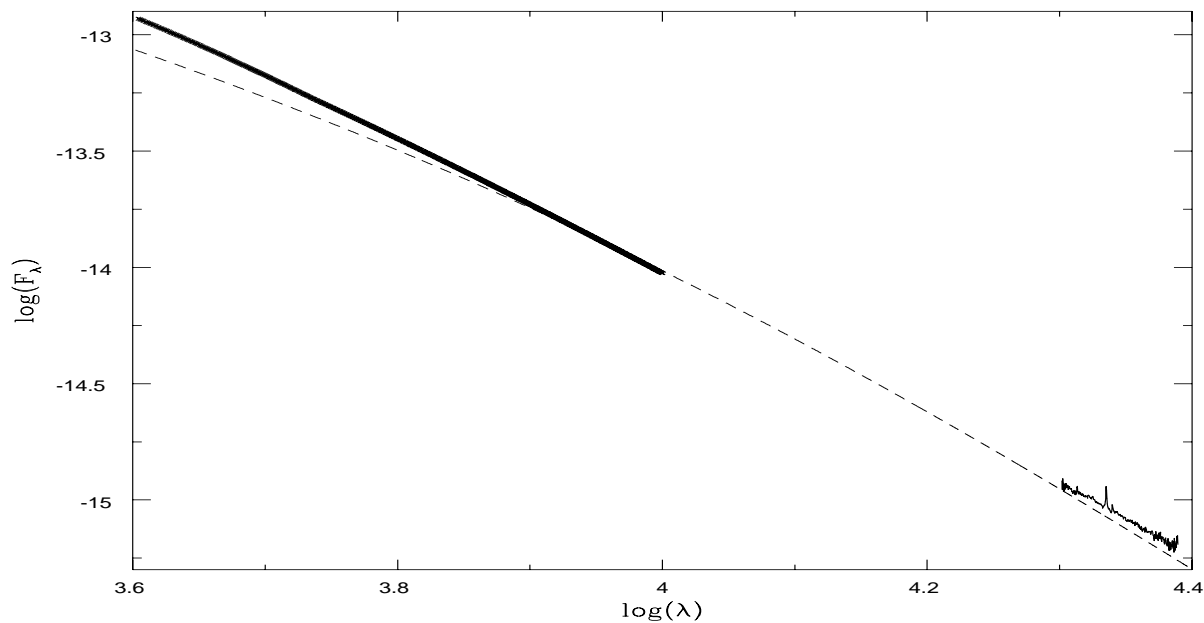}}}
\end{center}
\onecolumn{\caption{ The continuum of the observed optical (thick line) and $K$-band spectra of Sco X-1 are plotted together with the calculated spectrum ($F_{\lambda}$) of a steady state accretion disc with $T_{out}$ = 4200 K (dashed line) for comparison.  The parameters used for calculating the temperature were an outer disc radius of $6.3\times10^{10}$ cm (Vrtilek \etal 1991), $L_{X}$ = $3\times10^{37}$\ergs (Bradt \& McClintock 1983), and the inner radius of the disc at the last stable orbit around the neutron star.  The simulated disc spectrum has been arbitrarily normalized to match the flux of the optical spectrum at 0.9$\mu$m.}}
\end{figure} 

\begin{figure} 
\begin{center}
\rotatebox{90}{
\resizebox{0.5\textwidth}{0.7\textheight}{\includegraphics{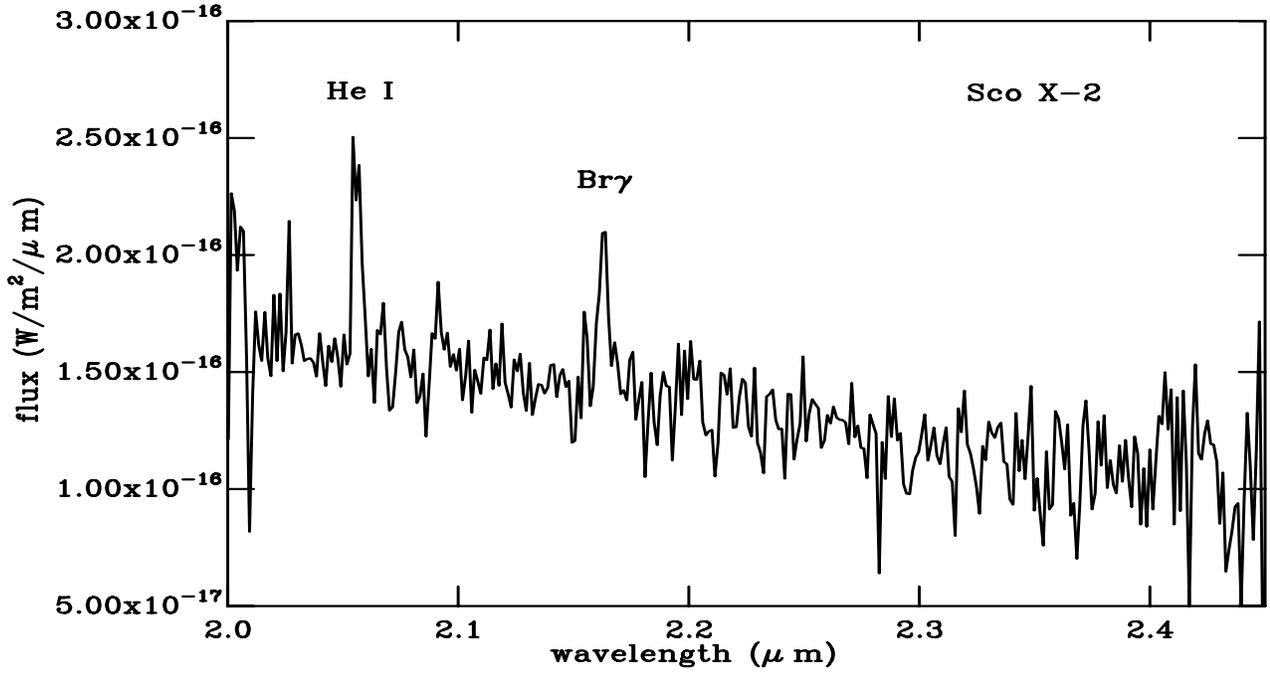}}}
\end{center}
\onecolumn{\caption{ $K$-band spectrum of the GBS Sco X-2 ($K\simeq$~14.6).  Br $\gamma$ and He I emission lines are clearly visible, confirming the identity of the IR counterpart to the X-ray source. }}
\end{figure} 

\begin{figure} 
\begin{center}
\rotatebox{90}{
\resizebox{0.5\textwidth}{0.7\textheight}{\includegraphics{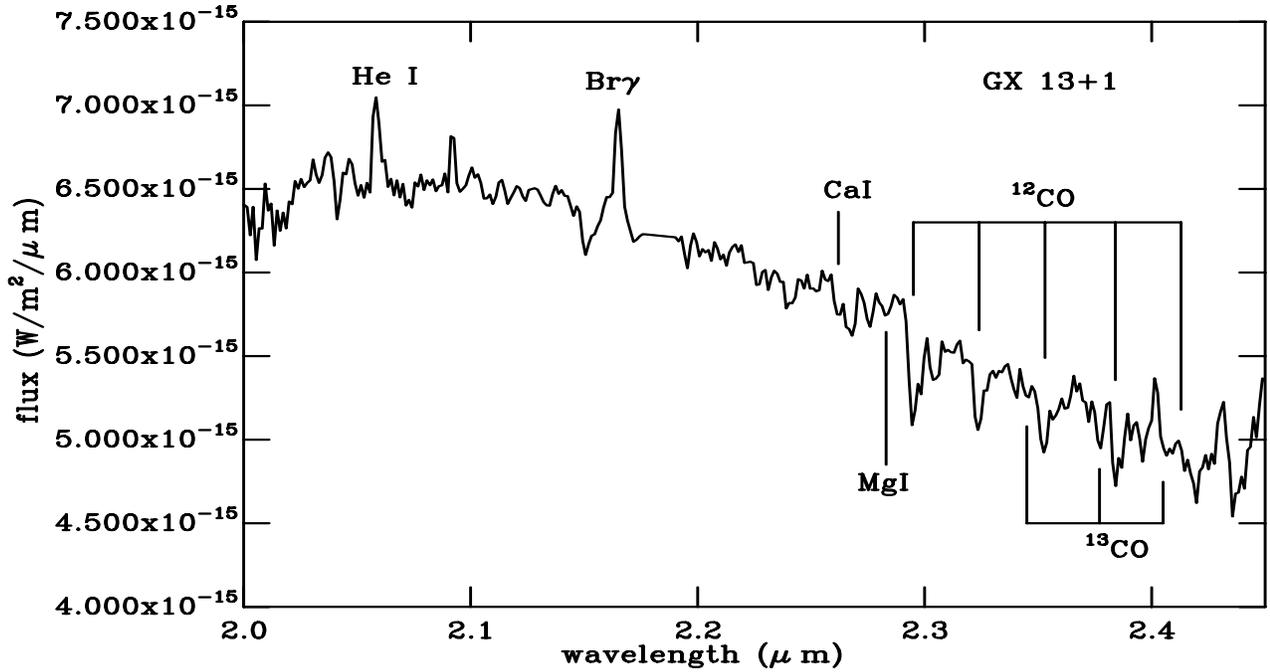}}}
\end{center}
\onecolumn{\caption{ $K$-band spectrum of the GBS GX13+1 ($K\simeq$~12.0).  Visible features include Br $\gamma$ and He I in emission and absorption bands of $^{12}$CO and $^{13}$CO, and with marginal detections of CaI and MgI.  The Br $\gamma$ line has a P Cygni profile with a strong absorption component.}}
\end{figure} 

\begin{figure} 
\begin{center}
\rotatebox{90}{
\resizebox{0.5\textwidth}{0.65\textheight}{\includegraphics{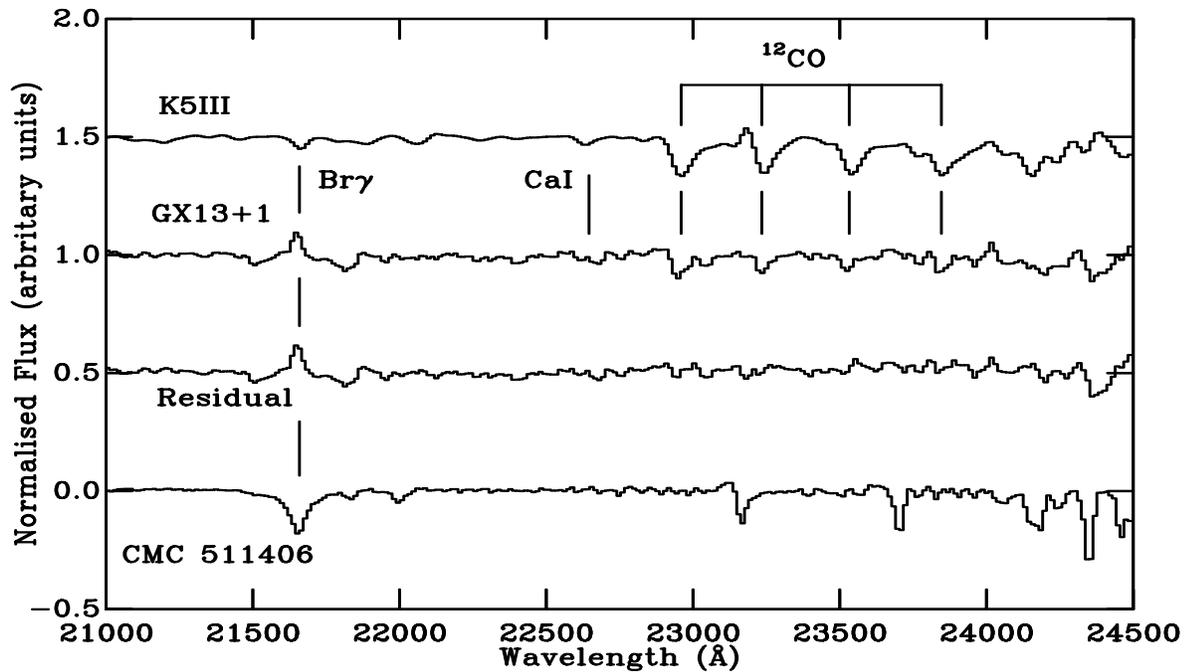}}}
\end{center}
\onecolumn{\caption{ This figure shows the spectrum of GX13+1 and that of the template K5$\sc iii$ star, as well as the residual after optimal subtraction.  The region including the first four $^{12}$CO bandheads was used to perform the subtraction.  Also shown is the A-type star used for removal of the telluric absorption features.  The flux level has been normalized and shifted for ease of plotting.}}
\end{figure} 

\begin{figure} 
\begin{center}
\rotatebox{90}{
\resizebox{0.5\textwidth}{0.7\textheight}{\includegraphics{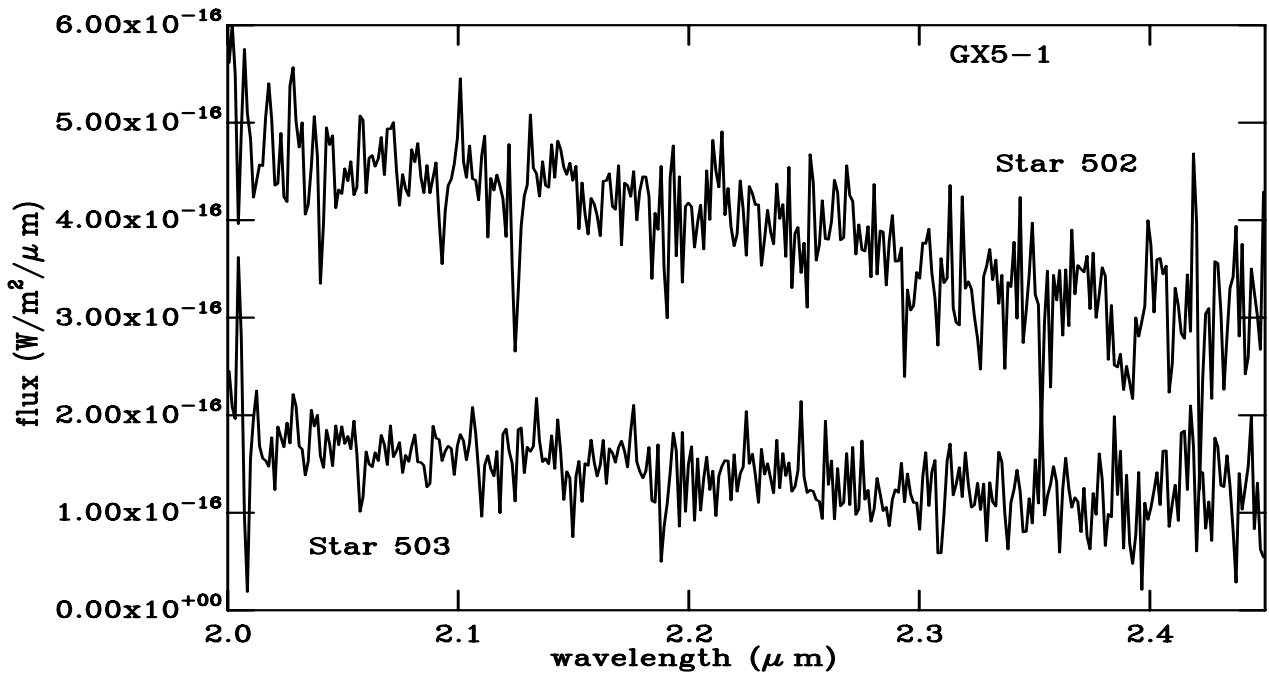}}}
\end{center}
\onecolumn{\caption{ The spectra of two candidates for the counterpart to GX5-1, stars 502 (top) and 503 (NCL91).}}
\end{figure} 

\begin{figure} 
\begin{center}
\rotatebox{-90}{
\resizebox{0.55\textwidth}{0.75\textheight}{\includegraphics{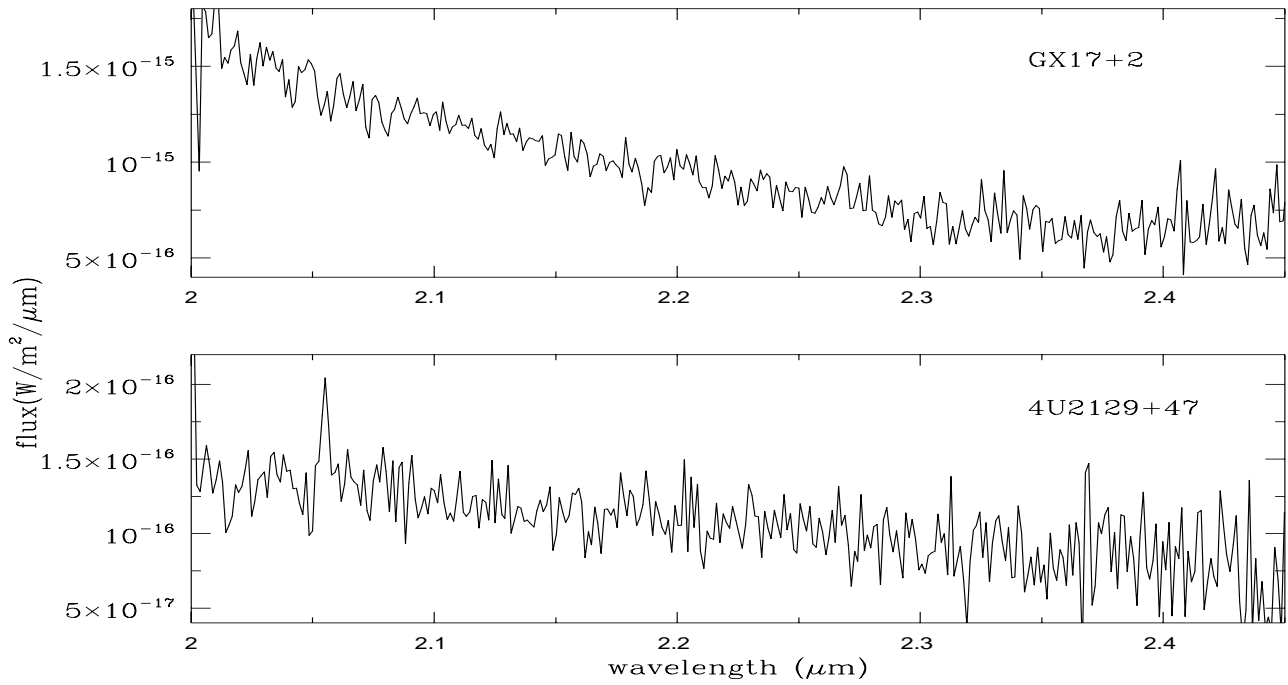}}}
\end{center}
\onecolumn{\caption{ The top panel is the spectrum of ``star TR'', the G star at the position of GX17+2.  The lower panel is the spectrum of the F star at the posision of 4U2129+47.  (The ``feature'' at 2.05 $\mu$m is a residual of the telluric absorption line removal.) }}
\end{figure} 

\end{document}